\documentstyle[preprint,prd,aps,floats,eqsecnum,epsf]{revtex}
\newcommand{\edc}{\end{document}}
\newcommand{\bb} {\begin{thebibliography}}
\newcommand{\eb}{\end{thebibliography}}

\newcommand{\bc}{\begin{center}}
\newcommand{\ec}{\end{center}}

\newcommand{\be}{\begin{equation}}
\newcommand{\ee}{\end{equation}}
\newcommand{\ba}{\begin{array}{l}   }
\newcommand{\lab}[1]{\label{#1}}
\newcommand{\ea}{\end{array}}

\newcommand{\dsfrac}{\displaystyle\frac}
\newcommand{\ds} {\displaystyle}

\newcommand{\re}[1]{(\ref{#1})}
\newcommand{\bra}[1]{\langle{#1}\vert}

\newcommand {\braket}[2]{ \bra{#1}{#2}\rangle }

\newcommand{\ci}{\cite}

\newcommand{\nwl}{\\[1mm]}

\newcommand{\hash}{\hbar }
\newcommand{\hinv}{{  \frac{ 1} {\hbar}     } }
\newcommand{\ejdj} {\ds{\exp\{  {\hash}j{\cd}j/2 \} } }
\newcommand{\ejgj} {\ds{\exp{\{\hash}jGj/2 \}} }
\newcommand{\intdx} {\ds{\int d^4x   } }
\newcommand{\intdf} {\ds{\int  \cal{D} \phi } }
\newcommand{\nolnol}{\braket{0}{0}}
\newcommand{{\cl}}{\cal {L}}

\newcommand{\veff}{V_{\mbox{eff}}}
\newcommand{\cd}{{\cal {D}}}
\newcommand{\dd}{\partial}
\newcommand{\dxx}{{\cd_{xx}}}
\newcommand{\tzjb}{\tilde Z[j,B]}
\newcommand{\twjb}{\tilde W[j,B]}
\newcommand{\tgjb}{\tilde \Gamma[\tilde\phi_0,B]}
\newcommand{ \delj }[1] {  \dsfrac{\delta #1}{\delta j(x)} }
\newcommand{\half}{\frac{1}{2}}
\newcommand{\mnol}{m_{0}^{2}}
\newcommand{\lamnol}{\lambda_0}
\newcommand{\omnol}{\Omega_{0}^{2}}
\newcommand{\om}{\Omega^2}

\newcommand{\inolm}{I_0(\Omega)}
\newcommand{\ibirm}{I_1(\Omega)}
\newcommand{\ikim}{I_2(\Omega)}
\newcommand{\iuchm}{I_3(\Omega)}
\newcommand{\iturm}{I_4(\Omega)}

\newcommand{\hata}[1]{\hat{A}^{(#1)} }
\newcommand{\Gxx}{{G_{xx}}}
\newcommand{\eps}{\varepsilon}
\newcommand{\phiz}{\phi_0}

\newcommand{\barmu}{\bar\mu}
\newcommand{\barphiz}{\bar\phiz}
\newcommand{\tochka}{\, .}
\newcommand{\vergul}{\, ,}
\tightenlines
\begin{document}
\draft
\title{\Large\bf {Optimized post Gaussian approximation in the background field method.}\\
%(\today)
}
\author{
 A. Rakhimov \thanks
{Permanant address: Institute of Nuclear Physics, Tashkent,
    Uzbekistan (CIS)}
    and
 Jae Hyung Yee \thanks{E-mail: jhyee@phya.yonsei.ac.kr}
 }
\address{
Institute of Physics and Applied Physics, Younsei University, Seoul, Korea\\}
\maketitle
\begin{abstract}
\medskip
\medskip
\medskip

We have extended the variational perturbative theory based
on the back ground field method to include  the optimized expansion of Okopinska and the post Gaussian effective
potential of Stansu and Stevenson. This new method provides much simpler way to compute
the correction terms to the Gausssian effective action (or potential). We have also
renormalized the effective potential in $3+1$ dimensions by introducing
 appropriate counter terms in the 
lagrangian.

\end{abstract}
\medskip
\medskip
%\pacs{PACS number(s):11.10.Wx, 11.15Bt, 13.85Fb}
%\keywords{finite temperature, scattering, decay }
\newpage

\section{Introduction}

The effective action and the effective potential have been used effectively in studying  various aspects
of  quantum field theories. Nonperturbative approximation methods for calculating them have been used
to extract essential features of nonlinear quantum field theories, 
such as properties of the ground state and information about  possible
phase transition of the system.

One of the most convenient nonperturbative approximations to the effective potential is
the  Gaussian variational approximation proposed more than  20 years ago \ci{hatfield}.
 The corresponding  Gaussian effective potential (GEP)  is known to contain one loop,
 sum of all daisy and superdaisy graphs of perturbation theory \ci{amelino} and  leading order in  $ 1/N$
expansion.

Lately some procedures to compute the correction terms to the GEP were proposed \ci{okop,stev42,cea55,yee}.
In our previous paper \ci{yee} we have developed a
 systematic method of the perturbative expansion around the Gaussian effective
action based on the background field method (BFM).
Although  BFM provides one with the simplest method of  computing the perturbative corrections to the GEP,
one can   extend the method to give a better approximation. 
The point is that in refs. \ci{cea55,yee}  the effective mass  $(\Omega)$  is
 fixed by the Gaussian 
gap equation, whereas 
 $\Omega$  is optimized at each stage of approximation  in refs.  \ci{okop,stev42}.

The main purpose of the present article is to extend BFM to include various
other post Gaussian methods developed so far.
 We shall  develop an improved post Gaussian effective potential (PGEP)
method  using BFM,
where $\Omega$ is not fixed preliminary.

The paper is organized as follows. In Sec. II we briefly present
the basic BFM  formulas, which will be necessary for  further
developments. In Sec. III we derive PGEP  using BFM
for the $\phi^4$ theory,  and we show that with BFM the computation of the higher order
corrections is very much simplified compared to other methods. In
Sec. IV we consider the renormalization procedure.
The summary and conclusions are  given in Sec. V.

\section{Background Field Method   for scalar fields}

For  readers convenience we outline here the basic formulas of
BFM. It is well known that, BFM is the most effective way of
doing the so called loop expansion \ci{abbot} 
 allowing us to study one - particle irreducible  (1PI) Feynman
diagrams.
The generating functional for disconnected graphs \footnote {Here and
below we use integral convention, e.g.
$j\phi\equiv\intdx j(x)\phi(x) \;$.}
is defined by
\be
Z[j]=\intdf \ds \exp\{i(S[\phi]+j\phi\}=\nolnol^j
\lab{bfm2.1}
\ee
where $j(x)$  is the external source and $S[\phi]$ is the classical action:
\be
S[\phi]=\intdx {\cl} [\phi(x),\dd_\mu\phi(x)] .
\lab{bfm2.2}
\ee
In BFM an analogous quantity, in which the classical action is written as a function of the field $\phi$
plus an arbitrary background field $B$, is defined by
\be
\tzjb=\intdf \ds \exp\{i(S[\phi+B]+j\phi\}   ,
\lab{bfm2.3}
\ee
where $  \tzjb$ depends both on the conventional source $j$ and on the  background field $B$.
Also by analogy with the conventional generator of connected graphs,
\be
W[j]=-i\ln Z[j]  ,
\lab{bfm2.4}
\ee
we define a new generating functional for connected
Green's functions,
\be
\twjb=-i\ln \tzjb  .
\lab{bfm2.5}
\ee

In the conventional approach the vacuum expectation value of the field operator in the presence  of
external source is defined by
\be
\phi_0(x)\equiv\langle\phi(x)\rangle^j=\delj{W[j]}\tochka
\lab{bfm2.6}
\ee
In BFM it is defined as
\be
\tilde\phi_0=\delj{\twjb} \vergul
\lab{bfm2.7}
\ee
 and the  background field effective action is defined by
\be
\tgjb =\twjb-j\tilde\phi_0 \tochka
\lab{bfm2.8}
\ee
Now it can be easily shown \ci{abbot} that the conventional  effective action can  be  calculated by the following equation:
\be
\Gamma[\phi_0] =\tilde \Gamma[0,B] \tochka
\lab{bfm2.9}
\ee
The background field effective action $\tgjb$ is just a conventional effective action computed in the presence of
 background field  $B$. It therefore consists of all $1PI$
graphs contributing to Green's functions. Since $1PI$  Green's functions are generated by taking derivatives of $\tgjb$
with respect to $\tilde\phi_0$ it  would generate $1PI$  Green's functions in the presence of the background field  $B$.
Now $\tilde \Gamma[0,B] $ has no dependence on  $\tilde\phi_0$, so it generates no graphs with external lines.
Instead, $\tilde \Gamma[0,B] $ is the sum all $1PI$ vacuum graphs in the
 presence of the  $B$ field.
The advantage of the  background field method is that many diagrams in the conventional
perturbation method are amalgamated into one diagram in BFM due to a new definition of the propagator function
\ci{abbot}. This method can be modified to extract the Gaussian part of the effective action out 
of the perturbative action part \ci{yee}. We will now show that BFM can be used to compute the
perturbative expansion  of the effective action around the Gaussian  effective action wihout fixing the effective mass.

\section{Post - Gaussian expansion of the effective potential}

For technical convenience we shall work throughout in the Euclidian formalism and start
with the Lagrangian  of the $\phi^4$ theory:
\be
{\cl}=\half \phi(-\dd^2+\mnol)\phi+\lamnol\phi^4 ,
\lab{bfm3.1}
\ee
where $m_0$ and $\lamnol$ are bare mass and bare coupling constant,  respectively.
Now following refs. \ci{okop,stev42} we rewrite \re{bfm3.1} in  an equal form  by introducing a new mass parameter
$\Omega_0$:
\be
{\cl}=\half \phi(-\dd^2+\omnol)\phi+\half(\mnol-\omnol)\phi^2 +\lamnol\phi^4 \tochka
\lab{bfm3.2}
\ee
The generating functional $\tzjb$ is then given by Eq. \re{bfm2.3}:
\be
\ba
\tzjb=\intdf \ds \exp\{-\hinv(S[\phi+B]+j\phi)\}=\intdf \ds \exp\{-\hinv(\intdx{\cl}[\phi+B]+j\phi)\}\vergul
\lab{bfm3.3}
\ea
\ee
where  ${\cl}(\phi+B)$ can be separated as follows:
\be
\ba
{\cl}(\phi+B)={\cl}_0(\phi)+{\cl}_1(\phi,B),
\nwl
{\cl}_0=\half \phi(-\dd^2+\omnol)\phi,
\nwl
{\cl}_1(\phi,B)=v_0(B)+v_1(B)\phi+v_2(B)\phi^2+v_3(B)\phi^3+v_4(B)\phi^4\vergul
\lab{bfm3.4}
\ea
\ee
with
\be
\ba
v_0(B)=\dsfrac{\mnol B^2}{2}+\lamnol B^4   ,  \quad  v_1(B)=B(\mnol+4\lamnol B^2),
\nwl
v_2(B)=\half(\mnol-\omnol)+6\lamnol B^2,  \;\;\;\;\; v_3(B)=4\lamnol B,
\nwl
v_4(B)=\lamnol \tochka
\lab{bfm3.5}
\ea
\ee
Now performing explicit Gaussian integration in Eq. \re{bfm3.3}  one obtains

\be
\ba
\tzjb=\ds \exp\{-\hinv\intdx{\cl}_{1}(\phi\rightarrow\delta/\delta j, B)\}\intdf \ds  \exp\{-(\hinv\intdx{\cl}_0(\phi)+j\phi) \}=
\nwl
=[\det {{\cd}}^{-1}]^{-\half}\exp\{-\hinv\intdx{\cl}_{1}(\phi\rightarrow\delta/\delta j,B)\}\ejdj \vergul
\lab{bfm3.6}
\ea
\ee
where
\be
{{\cd}}^{-1}_{xy}=(-\dd^2+\omnol)\delta_{xy}     \tochka
\lab{bfm3.7}
\ee

In accordance with our previous work  \ci{yee}, we introduce the  so called primed derivatives:

\be
\ba
(\dsfrac{\delta^2}{\delta j^{2}_{x}})'\equiv\hata{2}_x=\dsfrac{\delta^2}{\delta j^{2}_{x}}-\hash{\cd}_{xx}
\nwl
\nwl
(\dsfrac{\delta^3}{\delta j^{3}_{x}})'\equiv\hata{3}_x=\dsfrac{\delta^3}{\delta j^{3}_{x}}-3\hash^2{\cd}_{xx}R_x
\nwl
\nwl
(\dsfrac{\delta^4}{\delta j^{4}_{x}})'\equiv \hata{4}_x=\dsfrac{\delta^4}{\delta j^{4}_{x}}-
6\hash\dxx\dsfrac{\delta^2}{\delta j^{2}_{x}} +      3\hash^2\dxx^2 \vergul
\lab{bfm3.8}
\ea
\ee
where $R_x=\ds\int d^4y {\cd}_{xy}j(y)$, so that

\be
\hata{n}_{x}\ejdj=\hash^n R_{x}^{n}\ejdj\tochka
\lab{bfm3.9}
\ee
To isolate the Gaussian approximation we introduce another Green's function:
\be
\ba
\exp\{-\hinv(v_2\delta^2/\delta j^2+v_3\delta^3/\delta j^3+v_4\delta^4/\delta j^4)\}{\ejdj}
\nwl
=N_0 \exp\{\hinv a(B) \delta/\delta j\}\exp\{-\hinv(v_2 \hata{2}+v_3 \hata{3}+v_4 \hata{4})\}{\ejgj} \vergul
\lab{bfm3.10}
\ea
\ee
where $N_0$ , $a(B)$  and $G$ are  to be determined.
 By using the definition \re{bfm3.8} one finds the  following solution
 to the Eq.
\re{bfm3.10}:

\be
\ba
a(B)=-3v_3(B)\hash G_{xx},
\nwl
N_0(B)=(\det G^{-1})^{-1/2}  (\det {\cd}^{-1})^{1/2}
\exp\{-v_2 \Gxx+3v_4\hash G_{xx}^2\} \vergul
\lab{bfm3.11}
\ea
\ee
where $G_{xy}$ satisfies the  equation:
\be
G_{xy}^{-1}={\cd}^{-1}_{xy}+12\hash v_4\Gxx\delta_{xy}\tochka
\lab{bfm3.12}
\ee
The last equation looks rather simple in momentum space:
\be
\omnol=\om-12\hash\lamnol\inolm     ,\\
\lab{bfm3.13}
\ee
with
\be
\ba
{\cd}_{xy}=\ds\int\dsfrac{d^4p\exp[-ip(x-y)]}{(2\pi)^4(p^2+\omnol)},
\nwl
\nwl
G_{xy}=\ds\int\dsfrac{d^4p\exp[-ip(x-y)]}{(2\pi)^4(p^2+\om)},
\nwl
\nwl
\inolm=\ds\int\dsfrac{d^4p}{(2\pi)^4(p^2+\om)}\tochka
\lab{bfm3.14}
\ea
\ee
As to the determinant factors in Eq. \re{bfm3.11}, the first one can be written in a usual form:
\be
\ba
(\det G^{-1})^{-1/2}=\exp\{-\ibirm\}, \quad \quad  \ibirm=\half\ds\int\dsfrac{d^4p \ln(p^2+\om)}{(2\pi)^4},
\lab{bfm3.15}
\ea
\ee
  and the second one will be canceled by the term
$(\det {\cd}^{-1})^{-1/2}$ in Eq. \re{bfm3.6}.
Thus substituting Eqs. \re{bfm3.10}, \re{bfm3.11}, \re{bfm3.14} into \re{bfm3.6} we come to the final expression for
$ \tzjb$ and $\twjb$:
\be
\ba
\tzjb=\exp\{-\hinv[\hash\ibirm+v_0+\hash v_2\Gxx-3v_4\hash^2 G_{xx}^2]\} I_B\vergul
\nwl
\twjb/\hash=\ln\tzjb=-\ibirm-v_0/\hash-v_2 \Gxx+3v_4\hash G_{xx}^2+\ln I_B \vergul
\lab{bfm3.16}
\ea
\ee
with
\be
I_B=\exp\{-\hinv(v_1+3v_3\hash \Gxx)\delta/\delta j\}\exp\{-\hinv[v_2 \hata{2}+v_3 \hata{3}+v_4 \hata{4}]\}
{\ejgj}\tochka
\lab{bfm3.17}
\ee

The effective potential $V_{eff}$ is defined by
\be
V_{eff}(B)=-\dsfrac{\Gamma(B)}{\intdx}
\lab{bfm3.18}
\ee
 when  $ \Gamma(B)$ is independent of momenta. Using Eqs. \re{bfm2.5}-\re{bfm2.9} and \re{bfm3.16}-\re{bfm3.18} one can represent $V_{eff}$
 as a sum of the  GEP, $V_G$,  and a correction
$\Delta V_G$:
  \be
V_{eff}(B)=V_{G}(B)+\Delta V_G (B) \tochka
\lab{bfm3.19}
\ee
The explicit expression for $\Delta V_G$  will be given below. Here we note some points.

1) Due to the special construction of primed derivatives \re{bfm3.8}, the coefficient of $I_B$
in Eq. \re{bfm3.16}
 gives rise to the Gaussian effective action, so from Eqs.
   \re{bfm2.9}, \re{bfm3.16},  \re{bfm3.18} we immediately obtain the GEP:
  \be
\ba
V_{G}(B)=\hash\ibirm+v_0(B)+v_2\hash \Gxx-3v_4\hash^2G_{xx}^2
\nwl
\mbox{} =\hash\ibirm+v_0(B)+v_2\hash \inolm-3v_4\hash^2I_{0}^{2}(\Omega)
\nwl
\mbox{}=\hash\ibirm+\half\mnol B^2+\lamnol B^4+\hash \inolm[\half (\mnol-\omnol)+6\lamnol B^2]-
3\lamnol\hash^2I_{0}^{2}(\Omega)
\nwl
\mbox{}=\hash\ibirm+\half\mnol B^2+\lamnol B^4+\hash \inolm
[\half (\mnol-\om)+6\lamnol B^2]+3\lamnol\hash^2I_{0}^{2}(\Omega)\tochka
\lab{bfm3.20}
\ea
\ee
Here we used the explicit expressions for $v_0, v_2, v_4$ given by Eqs. \re{bfm3.5} and  \re{bfm3.13}.

2) The correction to the GEP,  that is, $\Delta V_G$,   in Eq. \re{bfm3.19} is  given by
% the logarithmic term in Eq.
%   \re{bfm3.16},  i.e.
 $\ln I_B$. As it was explained in \ci{yee} the linear term in the exponent in Eq. \re{bfm3.17}
can be omitted, since it does not contribute to the effective potential. Thus the corrections to the GEP will be given
by
 \be
\ba
\Delta V_G (B)=-\hash \ln I_B=-\hash\ln\{\exp[-\frac{\delta}{\hash}
(v_2 \hata{2}+v_3 \hata{3}+v_4 \hata{4})]\ejgj\}\ds\vert_{j=0}
\nwl
\nwl
=
-\hash \ln \{1- \dsfrac {\delta  (v_2 \hata{2}+v_3 \hata{3}+v_4 \hata{4})\exp[\hash jGj/2] }
 {\hash}\ds\vert_{j=0}+
\nwl
\nwl
+\dsfrac{\delta^2 (v_2 \hata{2}+v_3 \hata{3}+v_4 \hata{4})^2  \exp[\hash jGj/2]     }   {2\hash^2}    \ds\vert_{j=0}-
\nwl
\nwl
-\dsfrac{\delta^3 (v_2 \hata{2}+v_3 \hata{3}+v_4 \hata{4})^3  \exp[\hash jGj/2]     }   {6\hash^3}    \ds\vert_{j=0}
+\dots\}
\nwl
\equiv  \delta\Delta V_{G}^{(1)} (B)+
\delta^2\Delta V_{G}^{(2)} (B)+\delta^3\Delta V_{G}^{(3)} (B)+\dots  \tochka
\lab{bfm3.21}
\ea
\ee
Here we have introduced an auxiliary expansion parameter $\delta$ ( $\delta=1$).

The first order term     $ \Delta V_{G}^{(1)} (B)$
in  this equation will not contribute to the effective potential,  i.e.,
 $ \Delta V_{G}^{(1)} (B)=0$,   due to the relation,
$\hata{n}_{x}\ejgj\vert_{j=0}=\hash^n R_{x}^{n}\ejgj\vert_{j=0}=0$ (see Eq. \re{bfm3.9}).

3) As to the next term of the Eq. \re{bfm3.21} of order $\delta^2$
 the explicit calculations show that, only diagonal terms in the  expression
$(v_2 \hata{2}+v_3 \hata{3}+v_4 \hata{4})^2$ give a nonvanishing  result at $j=0$:
 \be
\ba
 (\hata{2})^2\ejgj \ds\vert_{j=0}=2!\hash^2\int d^4yG_{xy}^2
\nwl
 (\hata{3})^2\ejgj \ds\vert_{j=0}=3!\hash^3\int d^4yG_{xy}^3
\nwl
 (\hata{4})^2\ejgj \ds\vert_{j=0}=4!\hash^4\int d^4yG_{xy}^4 \tochka
\lab{bfm3.22}
\ea
\ee
Therefore,  to   $\delta^2$  order  we obtain the  following  expression
 for the first order correction to the  Gaussian effective potential,
\be
\Delta V_{G}^{(2)}(B)=-\dsfrac{ \hash }{2}
 [v_{2}^{2}\ikim+\hash v_{3}^{2}\iuchm+\hash^2 v_{4}^{2}\iturm   ]  \vergul
\lab{bfm3.23}
\ee
where in accordance with ref. \ci{stev42}, the following integrals are introduced:
\be
\ba
\dsfrac{\ikim}{2!}\equiv \int d^4yG_{xy}^2=\ds\int\dsfrac{d^4 p\; G^{2}(p)}{(2\pi)^4} ,
\nwl
\nwl
\dsfrac{\iuchm}{3!}\equiv \int d^4yG_{xy}^3=\ds\int\dsfrac{d^4 p d^4 q\;
 G(p) G(q)G(p+q) }{(2\pi)^8} ,
\nwl
\nwl
\dsfrac{\iturm}{4!}\equiv \int d^4yG_{xy}^4=
\ds\int\dsfrac{d^4 p d^4 q d^4 k \;G(p) G(q)G(k)G(p+q+k) }{(2\pi)^{12}}  ,
\nwl
 G(p)=1/(p^2+\om)\tochka
\lab{bfm3.24}
\ea
\ee

4) The expressions like $\{\hat {A}^{(m)}_{x_1}\hat {A}^{(n)}_{x_2}\dots \} \ejgj\ds\vert_{j=0}$
can be calculated analytically by using Mathemathica. The calculations show the
following result in $\delta^3$ order
\footnote{All other combinations vanish.}:
\be
\ba
\hat {A}^{(2)}_{x}\hat {A}^{(2)}_{y}\hat {A}^{(2)}_{z} \ejgj\ds\vert_{j=0}
=8G_{xy}G_{yz}G_{zx}  \vergul \;\;
\nwl
\hat {A}^{(2)}_{x}\hat {A}^{(2)}_{y}\hat {A}^{(4)}_{z} \ejgj\ds\vert_{j=0}
=24G_{xy}^{2}G_{yz}^{2} \vergul
\nwl
\hat {A}^{(2)}_{x}\hat {A}^{(3)}_{y}\hat {A}^{(3)}_{z} \ejgj\ds\vert_{j=0}
=36G_{xy}G_{yz}^{2}G_{zx}     \vergul \;\;
\nwl
\hat {A}^{(2)}_{x}\hat {A}^{(4)}_{y}\hat {A}^{(4)}_{z} \ejgj\ds\vert_{j=0}
=192G_{xy}  G_{yz}^{3}G_{zx} \vergul
\nwl
\hat {A}^{(3)}_{x}\hat {A}^{(3)}_{y}\hat {A}^{(4)}_{z} \ejgj\ds\vert_{j=0}
=216G_{xy}^{2}G_{yz}G_{zx}^{2}     \vergul \;\;
\nwl
\hat {A}^{(4)}_{x}\hat {A}^{(4)}_{y}\hat {A}^{(4)}_{z} \ejgj\ds\vert_{j=0}
=1728G_{xy}^{2}  G_{yz}^{2}G_{zx}^{2}  \; .
\lab{bfm3.25}
\ea
\ee
Using these equations in \re{bfm3.21} we get the second order  correction  ($\delta^3$
 terms) to the GEP,
\be
\ba

\Delta V_{G}^{(3)} (B)=\frac{4}{3}v_{2}^{3}G_{xy}G_{yz}G_{zx}+12v_{2}^{2}v_4 G_{xy}^{2}G_{yz}^{2}
+18v_{2}v_{3}^{2}G_{xy}G_{yz}^{2}G_{zx}+
\nwl
 +96v_{2}v_{4}^{2}G_{xy}  G_{yz}^{3}G_{zx}+108v_{3}^{2}v_4 G_{xy}^{2}G_{yz}G_{zx}^{2}+
288 v_{4}^{3}G_{xy}^{2}  G_{yz}^{2}G_{zx}^{2}   \tochka
\lab{bfm3.26}
\ea
\ee
 We see that, correction to the GEP
in $\delta^2$,        i.e. ,   $\Delta V_{G}^{(2)} (B)$ consists of only three Feynman diagrams shown in
Fig. 1 (a), which we call  BFM  diagrams. In Fig.1(a) quadratic vertex (marked as
   $\diamondsuit$ ) represents
\be
  v_2(B)=\half(\mnol-\omnol)+6\lamnol B^2,
\lab{bfm3.27}
\ee
On the other hand,  $\Delta V_{G}^{(2)} (B)$  includes five  diagrams found in ref. \ci{stev42}.
These  two additional diagrams
 may be obtained by using \re{bfm3.13} in \re{bfm3.27} and further in \re{bfm3.23}. As a
 result one gets all  diagrams
of ref.  \ci{stev42} shown in Fig. 1b.
One can see that the  diagrams of ref.  \ci{stev42} can be  obtained from the
BFM diagrams  by attaching a ring diagram with
$\inolm=\int d^4p/(2\pi)^4(p^2+\om)$
 to each $v_2$ vertex of a BFM diagram.

 Six BFM  diagrams in $\delta^3$ order  given by Eq. \re{bfm3.26},
 are presented in Fig. 2a. Now using Eqs. \re{bfm3.13} and  \re{bfm3.27}
 in \re{bfm3.26}, i.e. , by  attaching a ring diagram , denoted by a small circle
in Fig. 2a to each $v_2$ vertex,
one gets seven additional diagrams as is  shown in Fig.2b. After this procedure $\Delta V_{G}^{(3)} (B)$
will be the same as the one given by Okopinska \ci{okop}, represented by  13 diagrams.
 One may conclude that, the application of BFM technique
to study corrections to the GEP makes the work much easier than in the  formalism used in refs. \ci{stev42,okop}.
%The present method gives several BFM  diagrams and the application of Eqs. \re{bfm3.13}, \re{bfm3.27}
%gives some more additional diagrams needed for actual practical calculations.
 To illustrate this  we represent in
Fig.3  nine BFM  diagrams including $v_2$ vertices in $\delta^4$ order.
 The  implementation of a ring diagram, through Eqs.
\re{bfm3.13} and  \re{bfm3.27} gives  additional 18  diagrams.

The effective potential in  Eq. \re{bfm3.23} is exactly the same as that  given in refs. \ci{cea55,yee},
except that  the parameter $\Omega$ in Eq. \re{bfm3.23} is not fixed  by
the Gaussian part. In fact in accordance with the principle of "minimal sensitivity" $\Omega$ will be fixed by
$dV_{eff}/d\Omega=0$.

\section{Regularization and renormalization}

It is well known that, the renormalization of any field theory can be achieved
by introducing appropriate counter terms to the Lagrangian. In this section we apply 
this procedure to evaluate $V_{eff}$.
 To do this we rewrite the lagrangian \re{bfm3.1} including counter terms \ci{ramosprd60,chiku}:
 \be 
{\cl}=\half \phi(-\dd^2+m^2)\phi+\lambda\phi^4 +\half Bm^2\phi^2+C\lambda\phi^4-Dm^4 
\lab{bfm4.1}
 \ee
where $m$ and $ \lambda$ represent renormalized mass and renormalized 
coupling constant, respectively,
and make   the following mass  substitution: 
\be
 m^2=\om-\delta(\om-m^2)   \equiv\om-\delta\chi
 \lab{bfm4.2}
\ee
where $\delta$ is a superficial parameter, well known in the  delta expansion method.
 We then rewrite the Lagrangian in the same way as in $\delta $ expansion \ci{ramosprd60}
\be
{\cl}={\cl}_0+\delta({\cl}-{\cl}_0)\equiv \half \phi(-\dd^2+\Omega^2)\phi+{\cl}_{int}
 \lab{bfm4.3}
 \ee
where
\be
\ba
{\cl}_{\mbox{int}}=-\half\delta\chi \phi^2+\delta\lambda\phi^4+
{\cl}_{\mbox{cnt}}
\nwl
\nwl
{\cl}_{\mbox{cnt}}=\half \phi^2 B (\om-\delta\chi) + \lambda C\phi^4
-D(\om-\delta\chi)^2 \tochka
 \lab{bfm4.4}
\ea
\ee
 The idea  that the mass substitution \re{bfm4.2}  should 
be used not only in the standard mass term but also in the counter terms was first recognized in 
\ci{banerji} and further developed in \ci{chiku} where the nessesity of the counter term $D$
was also pointed out. 

We assume that the parameters B, C and D are expanded as
\be
\ba
B=B_1 \delta+B_2\delta+O(\delta^3)
\nwl
C=C_1\delta+C_2\delta^2+O(\delta^3)
\nwl
D=D_0+D_1\delta+D_2\delta^2+O(\delta^3) \tochka
\lab{bfm4.5}
\ea
\ee

Now using the method outlined in the previous sections one may get the
 following unrenormalized
 effective potential
up to the $\delta^2$ order
\be
\ba
V_{eff}(\phiz)=V_{G}(\phiz)+\Delta V_G (\phiz) \vergul
\nwl
V_{G}(\phiz)=\ibirm-D_0\Omega^4+\delta(v_0+v_2 \inolm+3v_4 I_{0}^{2}(\Omega))
\nwl
\Delta V_G (\phiz)= -\delta^2\{\half(v_2+6v_4\inolm)^2I_2(\Omega)+\half v_3^2I_3(\Omega)
+\half v_4^2I_4(\Omega)
\nwl
-(u_0+u_2\inolm+3u_4I_{0}^{2}(\Omega))\}\vergul
\lab{bfm4.6}
\ea
\ee
where
\be
\ba
v_0=\dsfrac{m^2\phiz^2}{2}+\lambda \phiz^4+ 
\dsfrac{B_1\om\phiz^2}{2}
+C_1\lambda \phiz^4+2D_0\om\chi-D_1\Omega^4  ,  
\nwl
 v_2=-\half\chi+6\lambda \phiz^2+\dsfrac{B_1\om}{2}
+ 6C_1\lambda \phiz^2 \vergul
\nwl
 v_3=4\lambda\phiz+4\lambda\phiz C_1\vergul
\quad\quad
v_4=\lambda+\lambda C_1 \vergul
\nwl
u_0=\dsfrac{(B_2\om-B_1\chi)\phiz^2}{2}+\lambda C_2 \phiz^4- 
D_2\Omega^4+2D_1\om\chi-D_0\chi^2 \vergul
\nwl
u_2=\dsfrac{(B_2\om-B_1\chi)}{2}+6\lambda C_2 \phiz^2 \vergul \quad\quad
%\nwl
 u_3=4\lambda\phiz C_2 \vergul
\quad\quad
u_4=\lambda C_2
\vergul
\lab{bfm4.7}
\ea
\ee
 and the divergent integrals $I_n(\Omega)$ calculated in the dimensional regularization
method in $3+1 $ dimensions  are brought to the Appendix.

We shall evaluate the regularized effective potential \footnote{More exactly  
$V_{eff}(\phiz,\Omega^2) -V_{eff}(0,m^2)$} order by order in powers of
$\delta$. 

It is easy to see that, the first divergent  term of the GEP ,
 coming from $I_1(\Omega)$ in Eq. \re{bfm4.6}
may be compensated by $D_0$ chosen as:
\be
\ba
{D_{0}} =  - {\displaystyle \frac {1}{32\pi ^{2}\varepsilon }}  
\tochka
\lab{bfm4.8}
\ea
\ee
As it is seen from Eq. \re {bfm4.1}, possible finite part of the counter term $D$ 
leads only to shifting the effective potential as a whole and may be neglected.
Using explicite expressions for  divergent integrals in Eqs.  \re{bfm4.6} and 
  \re{bfm4.7}the GEP is given by:
\be
\ba
V_{G}(\phiz)=\lambda  ({C_{1}} + 1) {\phi _{0}}^{4}+
M^{(2)}(\Omega){\phi _{0}}^{2}+
M^{(0)}_{{ln}}(\Omega)+M^{(0)}(\Omega)
\lab{bfm4.9}
\ea
\ee
where 
\be
\ba
 M^{(2)}(\Omega)= {\displaystyle 
\frac {3\Omega ^{2} \lambda ({C_{1}} + 1)
\mathrm{ln}({\displaystyle \frac {\Omega ^{2}}{\mu ^{2}}} )}{8\pi 
^{2}}}  
%\nwl
%\nwl
%\mbox{}
 + {\displaystyle \frac { 
 [3 \lambda({C_{1}} + 1
) (\gamma  - 1) + 4 B_1  \pi ^{2} ]
\Omega ^{2} + 4m^{2} \pi ^{2}   }{8\pi ^{2}}} 
\nwl
\quad\quad
 - {\displaystyle \frac {
3\Omega ^{2} \lambda  ({C_{1}} + 1) }{4\pi ^{2} 
\varepsilon }}  
\lab{bfm4.10}
\ea
\ee
and $M^{(0)}(\Omega)$ will be given below. From the first term of Eq. \re{bfm4.9}
  one may conlude that,  $C_1$ should be finite. The requrement of finiteness
 of $M^{(2)}(\Omega)$
in Eq. \re{bfm4.10} leads to : 
\be
\ba
{B_{1}} = {b_{0}} + {\displaystyle \frac { 3 \lambda 
 ({C_{1}} + 1)}{2\pi ^{2} \varepsilon }} 
\lab{bfm4.11}
\ea
\ee
where  $b_0$, as well as  $C_1$ are    finite constant to be determined by
suitablely chosen renormalization condition. Fortunately,  the choice of $B_1$ as in Eq.
\re{bfm4.11}  will cancel also the logarithmic pole like 
 ${ln(\Omega^2/\mu^2)}/{\varepsilon}$ coming 
from  $I_{0}^2 (\Omega)$ in Eq. \re{bfm4.6}.
The  term  $M^{(0)}_{{ln}}(\Omega)$  in Eq. \re{bfm4.9}  includes the logarithmic part,
\be
\ba
M^{(0)}_{{ln}}(\Omega)={\displaystyle \frac {3}{256}}  
{\displaystyle \frac {\lambda  ({C_{1}} + 1) (\Omega ^{4} 
\mathrm{ln}^{2}({\displaystyle \frac {\Omega ^{2}}{\mu ^{2}}} )
 - m^{4} \mathrm{ln}^{2}({\displaystyle \frac {m^{2}}{\mu ^{2}}}
 ))}{\pi ^{4}}}
\nwl
  + {\displaystyle \frac {1}{128\pi ^{4}}} \{
%\\
[(3 \lambda  ({C_{1}} + 1) (\gamma  - 1) + 4 {b_{0}} \pi ^{2
} - 2 \pi ^{2}) \Omega ^{4} + 4 m^{2} \pi ^{2} \Omega ^{2}]
 \mathrm{ln}({\displaystyle \frac {\Omega ^{2}}{\mu ^{2}}} ) 
\nwl
\mbox{} - [ m^{4} (3 \lambda  ({C_{1}} + 1) (\gamma  - 1)
 + 4 {b_{0}} \pi ^{2}) + 2 m^{4} \pi ^{2}] \mathrm{ln}(
{\displaystyle \frac {m^{2}}{\mu ^{2}}} )\}
 \vergul
\lab{bfm4.12}
\ea
\ee
which is now finite,  and the nonlogarithmic part 
\be
\ba
M^{(0)}(\Omega)={\displaystyle \frac {1}{256\pi ^{4}}} \{  (\Omega^{4} - 
m ^{4}) (\gamma  - 1) [3 \lambda  ({C_{1}} + 1) (\gamma
  - 1) + 8 {b_{0}} \pi ^{2}] + 2 \pi ^{2} (3 m^{4} + \Omega 
^{4})
\nwl
\\
 \mbox{} + 8 m^{2} \Omega ^{2} \pi ^{2} (\gamma  - 1) - 4 \pi
 ^{2} \gamma  (m^{4} + \Omega ^{4})
\nwl
\mbox{} + (m^{4} - \Omega ^{4}) [256 \pi ^{4} D_1
 + {\displaystyle \frac {16 {b_{0}} \pi ^{2
}}{\varepsilon }}  + {\displaystyle \frac {12 \lambda  ({C_{1}}
 + 1)}{\varepsilon ^{2}}} ]\} 
 \tochka
\lab{bfm4.13}
\ea
\ee
which becomes finite when
 $D_1$ is fixed as :
\be
{ {D}_{1}} =  - {\displaystyle \frac {1}{16}}  
{\displaystyle \frac {{b_{0}}}{\pi ^{2} \varepsilon }}  - 
{\displaystyle \frac {3}{64}}  {\displaystyle \frac {\lambda  (
{C_{1}} + 1)}{\pi ^{4} \varepsilon ^{2}}} \tochka
\lab{bfm4.14}
\ee

Now, after having removed  all the singularities   one 
comes to the following finite effective potential in $\delta$ order,
\be
\ba
 \veff^{(1)} (\phiz) =  {\displaystyle \frac {1}{2}}  m^{2}{\phi _{0}}^{2} \!  
   +\lambda  (C_{1} + 1) {\phi _{0}}^{4} + 
  \!
 \dsfrac{ \Omega ^{2} {\phi _{0}}^{2} { [}
 \! 3  \lambda  (C_{1} + 1) 
\mathrm{ln}(
\dsfrac { \Omega ^{2} } { \mu ^{2} } )
 + 
 3\lambda  (C_{1} + 1) + 4 b_{0} \pi ^{2})  \! 
{]} }{8\pi ^{2}} 
\nwl
\\
\mbox{} +
 \dsfrac {
3  \lambda  (C_{1} + 1) [\Omega ^{4} 
\mathrm{ln}^{2}(
{\displaystyle \frac {\Omega ^{2}}{\mu ^{2}}} )
 - m^{4} 
\mathrm{ln}^{2}({\displaystyle \frac {m^{2}}{\mu ^{2}}} )]
}
{256\pi ^{4}}
\nwl
\\
  + {\displaystyle \frac {1}{128 \pi ^{4}}} 
[ 
((3 \mathrm{\lambda  ({C_{1}} + 1)} + 4 {b_{0}} \pi ^{2} - 2 \pi ^{2}) \Omega 
^{4} + 4 m^{2} \pi ^{2} \Omega ^{2}) \mathrm{ln}(
{\displaystyle \frac {\Omega ^{2}}{\mu ^{2}}} ) 
\nwl
\\
\mbox{} - m^{4}(  (3 \mathrm{\lambda  ({C_{1}} + 1)}
 + 4 {b_{0}} \pi ^{2}) + 2
  \pi ^{2}) \mathrm{ln}({\displaystyle \frac {m^{2}}{\mu 
^{2}}} )] 
\mbox{} 
\nwl
\\
+ {\displaystyle \frac {1}{256\pi ^{4}}} 
[ 
 (\Omega^{4} - m ^{4}) (\gamma  - 1) (3 \mathrm{\lambda  ({C_{1}} + 1)} 
+ 8 {b
_{0}} \pi ^{2}) + 2 \pi ^{2} (3 m^{4} + \Omega ^{4})
\nwl
\\
 + 8 m^{
2} \Omega ^{2} \pi ^{2} (\gamma  - 1) 
\mbox{} - 4 \pi ^{2} \gamma  (m^{4} + \Omega ^{4})]\tochka
\lab{bfm4.15}
\ea
\ee
Note that, $\veff^{(1)} (\phiz)$ includes two finite constants
$b_0$ and
$ C_1$ to be determined by the suitably chosen renormalization conditions. 
If we employ the minimal subtraction (MS) 
scheme these constants may be neglected,  since in the MS scheme the
 counter terms are supposed to remove only
the singularities in $\eps$.
On the other hand, they  may be determined by  an intermediate  renormalization scheme 
\ci{itzykson}:
\be
\ba
\dsfrac{d^2 \veff (\phiz)}{d\phiz^2}\ds\vert_{\phiz=0}-m^{2}=0, \quad \quad
\dsfrac{1}{4!}\dsfrac{d^4 \veff (\phiz)}{d\phiz^4}\ds\vert_{\phiz=0}-\lambda=0,
\lab{bfm4.16}
\ea
\ee 
which is commonly used for fixing the parameters of the effective potential. In this case 
they may naturally depend on $ln(m^2/\mu^2)$.
From the first condition in Eq. \re{bfm4.16} the finite counter term
$b_0$ is easily determined as
\be
{b_{0}} =  - {\displaystyle \frac {3}{4}}  {\displaystyle 
\frac {\lambda  ({C_{1}} + 1) (\gamma  + \mathrm{ln}(
{\displaystyle \frac {m^{2}}{\mu ^{2}}} ) - 1)}{\pi ^{2}}} \vergul
\lab{bfm4.17}
\ee 
while  $C_1$  can be determined in principle from the second condition 
of Eq. \re{bfm4.16} which turned 
out to be a  nonlinear equation. We shall come back to this point later.

Now passing to the regularization of  $\veff$ in next order
 of $\delta $ we rewrite it  explicitely as:
\be
\ba
\Delta V_{G}(\phiz)=K^{(4)}(\Omega)\phiz^4+
[K^{(2)}_{ln}(\Omega)+K^{(2)}(\Omega)]\phiz^2+K^{(0)}_{ln}(\Omega)
+K^{(0)}(\Omega)
\lab{bfm4.18}
\ea
\ee   
where 
\be
\ba
K^{(4)}(\Omega)=  {\displaystyle 
\frac {\lambda  [9 \lambda  ({C_{1}} + 1)^{2} (\gamma  + 
\mathrm{ln}({\displaystyle \frac {\Omega ^{2}}{\mu ^{2}}} )) + 4
 {C_{2}} \pi ^{2}] }{4\pi ^{2}}}  -  {\displaystyle \frac {9\lambda ^{2} ({C_{1}} + 1)
^{2} }{2\pi ^{2} \varepsilon }}
\lab{bfm4.19}
\ea
\ee  
and $K^{(2)} (\Omega)$, $K^{(0)} (\Omega)$ will be given below.
 Finiteness of $K^{(4)} (\Omega)$
determines $C_2 $ as
\be
\ba
{C_{2}} = { {\tilde{C}}_{2}} + {\displaystyle \frac {
9  \lambda  ({C_{1}} + 1)^{2}}{2 \pi 
^{2} \varepsilon }}
\lab{bfm4.20}
\ea
\ee  
where $\tilde{C}_2$ is a finite constant.  The term $K^{(2)}_{ln}(\Omega)$ is given by:
\be
\ba
K^{(2)}_{ln}(\Omega)={\displaystyle \frac {45}{32}}  
{\displaystyle \frac {\lambda ^{2} \Omega ^{2} ({C_{1}} + 1)^{2
}  \mathrm{ln}^{2}({\displaystyle \frac {\Omega ^{2}}{\mu 
^{2}}} )}{\pi ^{4}}} 
\nwl
\nwl
- {\displaystyle \frac {3}{32 \pi ^{4}}} \lambda   \mathrm{ln}({\displaystyle \frac {\Omega ^{2}}{\mu ^{2}}} 
)\{3 \lambda  ({C_{1}} + 1)^{2} \Omega ^{2} \mathrm{ln}(
{\displaystyle \frac {m^{2}}{\mu ^{2}}} ) 
\nwl
 \\
\mbox{}+[(36\lambda  - 27\lambda  \gamma ){C_{1}}^{2} + (72 
\lambda  + 4\pi ^{2} - 54\lambda \gamma ) {C_{1}} + 4 \pi
 ^{2} + 36\lambda  - 4{ {\tilde{C}}_{2}}\pi ^{2} 
- 27\lambda \gamma]\Omega ^{2}
\nwl
\mbox{} - 4 \pi ^{2} ({C_{1}} + 1) m^{2}\} 
\mbox{} - {\displaystyle \frac {9}{8}}  {\displaystyle \frac {
\lambda ^{2}  \Omega ^{2} ({C_{1}} + 1)^{2} \mathrm{
ln}({\displaystyle \frac {\Omega ^{2}}{\mu ^{2}}} )}{\pi ^{4} 
\varepsilon }}  
\lab{bfm4.21}
\ea
\ee  
Here we see that there is a dangerous pole term propotional to
 $\mathrm{ln}(\Omega ^{2}/\mu ^{2} )/\varepsilon  $ which 
 can be removed only by the 
the  condition
\be
C_1=-1 \vergul
\lab{bfm4.22}
\ee 
since there is no infinite counter term in 
\re{bfm4.21} to compensate this pole.
Now the nonlogarithmic  term in \re{bfm4.18}, $K^{(2)}(\Omega)$,  is simply given by:
\be
\ba
K^{(2)}(\Omega)= {\displaystyle \frac {1}{8}}  {\displaystyle 
\frac {[4 {B_{2}} \pi ^{2} + 3 \lambda  { {\tilde{C}}_{2}} (
 - 1 + \gamma )]  \Omega ^{2}}{\pi ^{2}}}  - 
{\displaystyle \frac {3}{4}}  {\displaystyle \frac {\lambda  {
 {\tilde{C}}_{2}}  \Omega ^{2}}{\pi ^{2} \varepsilon }}
\lab{bfm4.23}
\ea
\ee  
and may be regularized by 
\be
\ba
{B_{2}} = {b_{2}} + {\displaystyle \frac {
3  \lambda  { {\tilde{C}}_{2}}}{2\pi ^{2} \varepsilon }} 
\lab{bfm4.24}
\ea
\ee  
where $b_2$ is a finite constant . 
One can see that the choice $C_1=-1$ will remove logarithmic 
  pole terms in
  $K^{(0)}_{ln}(\Omega)$ also :
\be
\ba
K^{(0)}_{ln}(\Omega)=  - {\displaystyle \frac {3}{256}}  
{\displaystyle \frac {\lambda  { {\tilde{C}}_{2}} [ m^{4} 
\mathrm{ln}^{2}({\displaystyle \frac {m^{2}}{\mu ^{2}}} ) - 
\Omega ^{4} \mathrm{ln}^{2}({\displaystyle \frac {\Omega ^{2}}{\mu 
^{2}}} ) ]}{\pi ^{4}}}
\nwl
\\
+{\displaystyle \frac {1}{128}}  
{\displaystyle \frac {(4 \Omega ^{4} {b_{2}} \pi ^{2} + 3 
\lambda  { {\tilde{C}}_{2}} \Omega ^{4} \gamma  - 3 \lambda  
{ {\tilde{C}}_{2}} \Omega ^{4} + 2 \Omega ^{4} \pi ^{2} + 2 m
^{4} \pi ^{2} - 4 \Omega ^{2} m^{2} \pi ^{2}) \mathrm{ln}(
{\displaystyle \frac {\Omega ^{2}}{\mu ^{2}}} )}{\pi ^{4}}} 
\nwl
\\
-{\displaystyle \frac {1}{128}}  
{\displaystyle \frac {m^{4} (4 {b_{2}} \pi ^{2} - 3 \lambda 
 { {\tilde{C}}_{2}} + 3 \lambda  { {\tilde{C}}_{2}} \gamma ) 
\mathrm{ln}({\displaystyle \frac {m^{2}}{\mu ^{2}}} )}{\pi ^{4}}}
\lab{bfm4.25}
\ea
\ee 
However, the nonlogariphmic term in Eq. \re{bfm4.18}, 
 $K^{(0)}(\Omega)$,  includes a simple pole term:
\be
\ba
K^{(0)}(\Omega)=
{\displaystyle \frac {(  m^{4} - \Omega ^{4}) [  - 8 \pi ^{2} 
( - 1 + \gamma ) {b_{2}} - 3 \lambda  ( - 1 + \gamma )^{2} {
 {\tilde{C}}_{2}} + 256{ {D}_{2}} \pi ^{4} ]  + 4 \pi ^{2
} \gamma  (m^{2} - \Omega ^{2})^{2}}{256\pi ^{4}}}
\nwl
\\
\mbox{} - {\displaystyle \frac {1}{16}}  {\displaystyle \frac {{
b_{2}} (\Omega ^{4} - m^{4})}{\pi ^{2} \varepsilon }}  - 
{\displaystyle \frac {3}{64}}  {\displaystyle \frac {\lambda  {
 {\tilde{C}}_{2}} (\Omega ^{4} - m^{4})}{\pi ^{4} \varepsilon ^{
2}}}  
\lab{bfm4.26}
\ea
\ee  
which can be made finite by the choice of $D_2$ as
\be
\ba
D_2= - {\displaystyle \frac {1}{16}}  
{\displaystyle \frac {{b_{2}}}{\pi ^{2} \varepsilon }}  - 
{\displaystyle \frac {3}{64}}  {\displaystyle \frac {\lambda  {
 {\tilde{C}}_{2}}}{\pi ^{4} \varepsilon ^{2}}} \tochka
\lab{bfm4.27}
\ea
\ee  

%\be
%\ba
%V_{eff}^{(2)}(\phiz)=V_G(\phiz)+\Delta V_G(\phiz)=\lambda  { {\tilde{C}}_{2}} \phiz ^{4} \\
%\nwl
%\mbox{} +  \{ \!   
%{\displaystyle \frac {3\Omega ^{2} \lambda  { {\tilde{C}}_{2}} 
%\mathrm{ln}({\displaystyle \frac {\Omega ^{2}}{\mu ^{2}}} )}{8\pi 
%^{2}}}  + {\displaystyle \frac {  (4
% \pi ^{2} m^{2} - 3 \Omega ^{2} \lambda  { {\tilde{C}}_{2}}
% + 4 \pi ^{2} \Omega ^{2} {b_{2}} + 3 \Omega ^{2} \lambda  
%{ {\tilde{C}}_{2}} \gamma )}{8\pi ^{2}}}  \! \}  \phiz ^{2}
 %\\
%\mbox{} -  {\displaystyle \frac {3
%\lambda  { {\tilde{C}}_{2}} (m^{4} \mathrm{ln}^{2}({\displaystyle 
%\frac {m^{2}}{\mu ^{2}}} ) - \Omega ^{4} \mathrm{ln}^{2}(
%{\displaystyle \frac {\Omega ^{2}}{\mu ^{2}}} ))}{256\pi ^{4}}} 
 %\\
%\mbox{} -   {\displaystyle \frac {
%m^{4} (3 \lambda  { {\tilde{C}}_{2}} \gamma  + 4 {b_{2}} \pi
 %^{2} + 2 \pi ^{2} - 3 \lambda  { {\tilde{C}}_{2}}) \mathrm{ln
%}({\displaystyle \frac {m^{2}}{\mu ^{2}}} )}{128\pi ^{4}}}
  %\\
%\mbox{} + {\displaystyle \frac {  
%(8 \Omega ^{4} {b_{2}} \pi ^{2} - 6 \lambda  { {\tilde{C}}_{2
%}} \Omega ^{4} + 4 m^{4} \pi ^{2} + 6 \lambda  { {\tilde{C}}
%_{2}} \Omega ^{4} \gamma ) \mathrm{ln}({\displaystyle \frac {
%\Omega ^{2}}{\mu ^{2}}} )}{256\pi ^{4}}}  
%\nwl
%\nwl
%+ {\displaystyle \frac {  ((\Omega ^{4} - m^{4}) ( - 1 + 
%\gamma ) (3 \lambda  { {\tilde{C}}_{2}} \gamma  + 8 {b_{2}} 
%\pi ^{2} - 3 \lambda  { {\tilde{C}}_{2}}) + 2 \pi ^{2} ( - 4 
%\Omega ^{2} m^{2} + \Omega ^{4} + 3 m^{4}))}{256\pi ^{4}}}  
%\tochka
%\lab{bfm4.28}
%\ea
%\ee

Now the effective potential has been successfully regularized up to $\delta^2$
order and  is given  by 
\be
\ba
\bar{V}^{(2)}_{eff}(\barphiz)={\displaystyle \frac {1}{256\pi ^{4}}} (256 \lambda  {
 {\tilde{C}}_{2}} \barphiz ^{4} \pi ^{4} + 32 \pi ^{2} (4 \pi ^{2
} (1 + {b_{2}} x) + 3 x \lambda  { {\tilde{C}}_{2}} (
\mathrm{ln}(x \barmu ) + \gamma  - 1)) \barphiz ^{2}
\nwl
\\
\mbox{} + 3 \lambda  { {\tilde{C}}_{2}} (x^{2} \mathrm{ln}(x 
\barmu )^{2} - \mathrm{ln}(\barmu )^{2}) + (6
\lambda  { {\tilde{C}}_{2}}(1 -\gamma )  - 4 \pi ^{2} (2 {b_{2}} + 1)) 
\mathrm{ln}(\barmu )
\nwl
\\
\mbox{} + ((6 \gamma  - 6) \lambda  x^{2} { {\tilde{C}}_{2}}
 + 4 \pi ^{2} (1 + 2 {b_{2}} x^{2})) \mathrm{ln}(x \barmu )
 + 3 \lambda  (\gamma  - 1)^{2} (x^{2} - 1) { {\tilde{C}}_{2}}
 \nwl
\\
\mbox{} + 2 (x - 1) \pi ^{2} (4 {b_{2}} (\gamma  - 1) (x + 
1) + x - 3))
% \left/ {\vrule height0.43em width0em depth0.43em}
 %\right. \!  \!  }
\lab{bfm4.28}
\ea
\ee
where we introduced dimensionless parameters $x$, $\barmu$ , $\barphiz$ defined by
\be
\Omega^2 = x m^2 ,  \quad  \mu^2 = m^2/\barmu \vergul 
\quad \quad \phiz  = m\barphiz  \vergul
\lab{bfm4.29}
\ee
and  presented  the effective potential in the units of $m^4$. 
The effective potential  includes two extra parameters  $\Omega$ and $\mu$. 
 In practical calculations 
the former may be determined by the principle of minimal sensitivity (PMS)  \ci{stev23}  or 
alternatively, by the fastest apparent convergence (FAC)  conditions
 ( see e.g. \ci{banerji}) while the latter
may be fixed by a renormalization scheme.

 Below we shall 
use  the PMS   which is natural for the renormalization scheme given in Eqs. \re{bfm4.16}.
In accordance with the PMS the optimum value of $\Omega$ will be given
by the equation, $d\veff/d\Omega=0$ , so, differentiating  \re{bfm4.28} with respect to 
$x$  leads to the following 'gap' equation:
 \be
\ba
16 \pi ^{2} x [4 {b_{2}} \pi ^{2} + 3 
\lambda  { {\tilde{C}}_{2}} \mathrm{ln}(x) + 3 \lambda  {
 {\tilde{C}}_{2}} \gamma  + 3 \lambda  { {\tilde{C}}_{2}} 
 {\mathrm{ln}(\barmu)}] \barphiz ^{2} 
\nwl
\\
\mbox{} + 3 \lambda  { {\tilde{C}}_{2}} x^{2} ( {\mathrm{ln}^{2}(\barmu)}
 + \mathrm{ln}^{2}(x))+ 2 \pi ^{2} [2 {b_{2}} x^{2} (2 \gamma  - 1) + (x
 - 1)^{2}] 
 \nwl
\\
\mbox{} + [ 6 \lambda  { {\tilde{C}}_{2}} x^{2}  {\mathrm{ln}(\barmu)} + 
(6 x^{2} \lambda  \gamma  - 3 x^{2} \lambda ) { {\tilde{C}}
_{2}} + 8 x^{2} {b_{2}} \pi ^{2} ] \mathrm{ln}(x)
\nwl
\\
\mbox{} + x^{2} (6 \lambda  { {\tilde{C}}_{2}} \gamma  - 3 
\lambda  { {\tilde{C}}_{2}} + 8 {b_{2}} \pi ^{2})  {\mathrm{ln}(\barmu)}
 + 3 \gamma  x^{2} \lambda  (\gamma  - 1) { {\tilde{C}}_{2}}=0
\lab{bfm4.30}
\ea
\ee 
which determines $\Omega$ as a function of $\phiz$. Note that, this dependence, or,
more exactly, $d\Omega/d\phiz$ should be explicitely used when calculating high
order total derivatives of $\veff$ with respect to $\phiz$, for example, 
in Eqs. \re{bfm4.16}. The first condition in Eqs. \re{bfm4.16} gives:
\be
\ba
3 \lambda  { {\tilde{C}}_{2}} ({\mathrm{ln}(\barmu)} - 1
 + \gamma ) + 4 {b_{2}} \pi ^{2}=0
\lab{bfm4.31}
\ea
\ee 
which determines $b_2$ as
\be
\ba
 {b_{2}} =  - {\displaystyle \frac {3}{4}}  {\displaystyle 
\frac {\lambda  { {\tilde{C}}_{2}} ( {\mathrm{ln}(\barmu)} - 1 + \gamma )
}{\pi ^{2}}} 
\lab{bfm4.32}
\ea
\ee 
and the second condition
\be
\ba
 [2 (\gamma  +  {\mathrm{ln}(\barmu)})^{2} - 4  {
\mathrm{ln}(\barmu)} + 4 - 4 \gamma ] { {\tilde{C}}_{2}} - 2 (\gamma  + 
 {\mathrm{ln}(\barmu)})^{2} - 1 + 4 \gamma  + 4  {\mathrm{ln}(\barmu)}=0
\lab{bfm4.33}
\ea
\ee 
 determines $\tilde{C}_2$ as:
\be
\ba
{ {\tilde{C}}_{2}} = {\displaystyle \frac {1}{2}}  
{\displaystyle \frac {2 (\gamma  +  {\mathrm{ln}(\barmu)})^{2} - 4 
 {\mathrm{ln}(\barmu)} - 4 \gamma  + 1}{(\gamma  +  {\mathrm{ln}(\barmu)})^{2} + 2
 - 2  {\mathrm{ln}(\barmu)} - 2 \gamma }} 
\lab{bfm4.34}\tochka
\ea
\ee 
Now substituting this into \re{bfm4.30} and considering the point $\phiz=0$,
$x=1$, (i.e. $\Omega=m$ at $\phiz=0$) we get :
\be
\ba
2 \lambda   {\mathrm{ln}^{4}(\barmu)} + 8 \lambda  (
 - 1 + \gamma )  {\mathrm{ln}^{3}(\barmu)} + \lambda  ( - 24 \gamma  + 
12 \gamma ^{2} + 11)  {\mathrm{ln}^{2}(\barmu)}
\nwl
\\
\mbox{} + 2 \lambda  ( - 1 + \gamma ) (2 \gamma  - 1) (2 
\gamma  - 3)  {\mathrm{ln}(\barmu)} + \lambda  (2 \gamma ^{2} - 4 
\gamma  + 1) ( - 1 + \gamma )^{2} =0\tochka
\lab{bfm4.35}
\ea
\ee
One can easily check that this equation is satisfied for 
 \be
\ba
 {\mathrm{ln}(\barmu)} =  {\mathrm{ln}(m^2/\mu^2)}=1 - \gamma  
\lab{bfm4.36}
\ea
\ee 
which leads to the following values for the finite  counter terms:
\be
\ba
b_2=0 \quad\vergul  \quad \tilde{C}_2=- \dsfrac{1}{2}
 \tochka
\lab{bfm4.37}
\ea
\ee 
Finally, the renormalized effective potential up to  $\delta^2$ order is given by:
\be
\ba
 \bar{V}^{(2)}_{eff}(\barphiz) =  - {\displaystyle \frac {1}{2}}  \lambda  \barphiz 
^{4} + {\displaystyle \frac {  [ - 
3 x \lambda  \mathrm{ln}(x) + 3 x \lambda  + 16 \pi ^{2}] 
\barphiz ^{2}}{32\pi ^{2}}} 
\nwl
 \\
\mbox{} -   {\displaystyle \frac {
8 \pi ^{2} (x - 1) - \mathrm{ln}(x) (3 x^{2} \lambda  + 8 
\pi ^{2})}{512\pi ^{4}}}  
\vergul
\lab{bfm4.38}
\ea
\ee 
with x  being the solution of the gap equation, 
\be
\ba
3 x \lambda  [\mathrm{ln}(x) + 1] [x 
\mathrm{ln}(x) + 16 \barphiz ^{2} \pi ^{2}] - 4 \pi ^{2} (x - 1)
^{2}=0
\lab{bfm4.39}
\ea
\ee 

Comparing Eqs. \re{bfm4.38}, \re{bfm4.39} with the results 
of  the "precarious" version of renormalization
 obtained by Stancu and Stevenson
( Eqs. (3.30) and (3.21) of  \ci{stev42}) , one can
see that their mathematical forms 
  are different. One of the reasons of this difference is that, in ref. \ci{stev42}
the finite part of two and three loop integrals  $I_3$ and $I_4$  were not given
 while in the present work we used the explicit form
of the finite part ( see Appendix).
The second reason may be hidden in the fact that, to make the effective potenial
finite the most authors use  the procedure "regularization after renormalization" while in the
present work we first regularized the potential, i.e. made it finite, and then applied the
renormalization conditions - Eqs. \re{bfm4.16}. 
Note that, the nessesity that the regularization  should be carried out before imposing
the gap equation was  pointed out also  in refs.  \ci{chiku,ramos65}.
 
\section{Conclusions}

We have developed a systematic method
for  computing the correction terms   to  the
Gaussian effective action. They are  given   by the expansion of:
\be
\exp\{-\frac{\delta}{\hash} [v_2 \hata{2}+v_3 \hata{3}+v_4 \hata{4}]\}{\ejgj}
\lab{bfm5.2}
\ee
in powers of $\delta$.
  Although our final  result
for unrenormalized PGEP is the same as that of refs. \ci{okop} and  \ci{stev42},
the method proposed in the
 present work has some advantages.

1)In refs.  \ci{okop,stev42} the expansion parameter $\delta$ is introduced directly into the Lagrangian
at the very beginning of calculations. Instead, by using BFM we preliminary extracted the
 Gaussian part
of the effective
potential and then introduced an expansion parameter $\delta$ in Eq. \re{bfm3.17}.
 This seems to be  a more natural way, since in accordance with general
 rules of quantum mechanics exponent of any
operator $\hat A$ should be understood as a Taylor expansion, i.e.:
$\exp(\hat A)=\{1+\delta \hat A+(\delta \hat A)^2/2!+(\delta \hat A)^3/3!+\dots\}\vert_{\delta=1}$.

2)In the present approach
Gaussian part
of the effective action is  extracted by
   introducing the  primed derivatives  \re{bfm3.8}.
We have shown that, in our method the computation of correction terms
to the GEP is greatly   simplified.
In fact, at each order of $\delta$
it gives  BFM diagrams (three in $\delta^2$ order and six  in $\delta^3$ order)
with a quadratic vertex $v_2=\half(\mnol-\omnol)+6\lamnol \phi^2  $ , cubic vertex $4\lambda_0\phi$ and quartic
vertex  $\lambda_0$. These diagrams do not include a ring diagram ($G_{xx}~I_0)   $
explicitly, which is  included  there   through the relation $\omnol=\om-12\hash\lamnol\inolm$,
giving rise to the  ones
(five in  $\delta^2$ order and 13  in $\delta^3$ order) appearing in the methods used in ref.s \ci{okop,stev42}.

3)In the present method it is easier to solve  equations \re{bfm2.7} -\re{bfm2.9}
 of BFM than similar equations in  the method used  in
\ci{stev42}.
In fact, in the  method of refs.   \ci{okop,stev42} one has to solve the equations
 $\Gamma[\phi]=W[J]-J\phi$, $\phi=\delta W/\delta J$ to define $\phi$ and $J$ at each order
of $\delta$. Although the corresponding solutions are simple in $\delta^2$ order
 \footnote{
Eq. (2.29)
 of ref. \ci{stev42}.
}
,  $J=\delta (v_1+3v_0I_0)+O(\delta^2)$, they are very complicated at higher orders.
On the other hand,  we have shown that due to the advantages
 of BFM one may simply put $J=0$ at each order of $\delta$.
Thus using the method proposed in the present paper  it will be  much simpler
to calculate higher  order corrections to GEP in general.

We have shown that the variational perturbation expansion of
 $\phi^4$ scalar theory  in four dimensions  can be successfully renormalized 
 by introducing appropriate counter terms in such a way that 
the whole Lagrangian has the same polynomial
 form as the bare  one.

\section*{Acknowledgments}
We thank  W. F. Lu and Prof. Chul Koo Kim    for useful discussions.
  A.M.R.            is
indebted to the Yonsei  University  for hospitality
 during his stay, where   the main part of
this work was  performed. This research was in part
 supported by BK21 project and in part by Korea Research Foundation under project
number KRF-2001-005-D20003 (JHY).
\newpage
\appendix
\bc
 {\bf Appendix}
\ec
\section*{Explicite expression for divergent integrals.}
\setcounter{section}{1}
%\label{AP.A}
%{\bf A. Explicite expression for divergent integrals.}\\
Here we bring explicit expressions for the divergent integrals defined as :
\small
\be
\ba
\inolm=\ds\int\dsfrac{d^4p}{(2\pi)^4(p^2+\om)}\vergul  \quad \quad  
\ibirm=\half\ds\int\dsfrac{d^4p \ln(p^2+\om)}{(2\pi)^4},
\nwl
\nwl
\dsfrac{\ikim}{2!}\equiv\ds\int\dsfrac{d^4 p\; G^{2}(p)}{(2\pi)^4} ,\quad \quad G(p)=1/(p^2+\om),
\nwl
\nwl
\dsfrac{\iuchm}{3!}\equiv \ds\int\dsfrac{d^4 p d^4 q\;
 G(p) G(q)G(p+q) }{(2\pi)^8} ,
\nwl
\nwl
\dsfrac{\iturm}{4!}\equiv \ds\int\dsfrac{d^4 p d^4 q d^4 k \;G(p) G(q)G(k)G(p+q+k) }{(2\pi)^{12}}  ,
 \ea
\lab{bfmAp.1}
\ee 
in four dimension. Integrals $\inolm,  \ibirm    $ and $\ikim$ are quite simple and
may be found  elsewhere \ci{stev32}.
\be
\ba
\inolm=  - {\displaystyle \frac {1}{8}}  
{\displaystyle \frac {\Omega ^{2}}{\pi ^{2} \varepsilon }}  + 
{\displaystyle \frac {  \Omega ^{2}
 [ - 1 + \gamma_{0}   + \mathrm{ln}({\displaystyle \frac {1}{4}}  
{\displaystyle \frac {\Omega ^{2}}{\mu ^{2} \pi ^{2}}} )]}{16\pi 
^{2}}} 
\nwl
\nwl
\ikim= {\displaystyle \frac {1}{4}}  
{\displaystyle \frac {1}{\pi ^{2} \varepsilon }}  - 
{\displaystyle \frac {1}{8}}  {\displaystyle \frac {\gamma_{0}  + 
\mathrm{ln}({\displaystyle \frac {1}{4}}  {\displaystyle \frac {
\Omega ^{2}}{\mu ^{2} \pi ^{2}}} )}{\pi ^{2}}}
\nwl
\nwl
\ibirm= I_{1}(m) + {\displaystyle 
\frac {1}{2}}  (\Omega ^{2} - m^{2}) {I_{0}}(m) - 
{\displaystyle \frac {1}{8}}  (\Omega ^{2} - m^{2})^{2} {
 {I}_{2}(m)} \nwl
\nwl
\mbox{} + {\displaystyle \frac {  m
^{4} [2  x^{2} \mathrm{ln}(x) - 
2  x + 2 
 - 3  (x - 1)^{2}]}{128\pi ^{2}}}
\ea
\lab{bfmA.2}
\ee 
where $x=\om/m^2$,  $\eps$ is defined as $d=4-\eps$ , d- space - time
dimension, $\gamma_{0}=0.5777215$ and  $\mu$ is an arbitrary constant with
mass dimension which usually appears in dimensional regularization
scheme. 

The two loop $I_3(\Omega)$ and three loop integrals $I_4(\Omega)$ have been
accurately calculated in refs. \ci{chung} including finite
parts:  
\be
\ba
{ {I}_{3}(\Omega)} =  - {\displaystyle \frac {3}{256\pi ^{4} }} \Omega ^{2
} \left( {\vrule height1.42em width0em depth1.42em} \right. \! 
 \! 24 {\displaystyle \frac {1}{\varepsilon ^{2}}}  + 
{\displaystyle \frac { - 24 \gamma_{0}  - 24 \mathrm{ln}(
{\displaystyle \frac {1}{4}}  {\displaystyle \frac {\Omega ^{2}
}{\pi  \mu ^{2}}} ) + 36}{\varepsilon }}  + 12 \mathrm{ln^2}({\displaystyle \frac {1
}{4}}  {\displaystyle \frac {\Omega ^{2}}{\pi  \mu ^{2}}} )         \nwl
\nwl
\mbox{} + (24 \gamma_{0}  - 36) \mathrm{ln}({\displaystyle \frac {1
}{4}}  {\displaystyle \frac {\Omega ^{2}}{\pi  \mu ^{2}}} ) + 
12 {\tilde{A}} - 36 \gamma_{0}  + 12 \gamma_{0} ^{2} + \pi ^{2} + 42
 \! \! \left. {\vrule height1.42em width0em depth1.42em} \right) 
 \nwl
\nwl
\nwl
{ {I}_{4}}(\Omega) = {\displaystyle \frac {3}{512\pi ^{6}}} \Omega ^{4}
 \left( {\vrule height1.49em width0em depth1.49em} \right. \! 
 \! 16 {\displaystyle \frac {1}{\varepsilon ^{3}}}  + 
{\displaystyle \frac {{\displaystyle \frac {92}{3}}  - 24 \gamma_{0}
  - 24 \mathrm{ln}({\displaystyle \frac {1}{4}}  
{\displaystyle \frac {\Omega ^{2}}{\pi  \mu ^{2}}} )}{
\varepsilon ^{2}}}  \nwl
\nwl
\mbox{} + {\displaystyle \frac {35 - 46 \gamma_{0}  + 18 \gamma_{0} ^{2
} + \pi ^{2} + (36 \gamma_{0}  - 46) \mathrm{ln}({\displaystyle 
\frac {1}{4}}  {\displaystyle \frac {\Omega ^{2}}{\pi  \mu ^{2}
}} ) + 18 \mathrm{ln}^{2}({\displaystyle \frac {1}{4}}  
{\displaystyle \frac {\Omega ^{2}}{\pi  \mu ^{2}}} )}{
\varepsilon }}  \nwl
\nwl
\mbox{} + \mathrm{ln}({\displaystyle \frac {1}{4}}  
{\displaystyle \frac {\Omega ^{2}}{\pi  \mu ^{2}}} ) ( - 
{\displaystyle \frac {105}{2}}  + 69 \gamma_{0}  - 27 \gamma_{0} ^{2}
 - {\displaystyle \frac {3}{2}}  \pi ^{2}) + \mathrm{ln}^{2}(
{\displaystyle \frac {1}{4}}  {\displaystyle \frac {\Omega ^{2}
}{\pi  \mu ^{2}}} ) ({\displaystyle \frac {69}{2}}  - 27 
\gamma_{0} )
%\nwl
\mbox{} - 9 \mathrm{ln}^{3}({\displaystyle \frac {1}{4}}  
{\displaystyle \frac {\Omega ^{2}}{\pi  \mu ^{2}}} ) 
 \! \! \left. {\vrule height1.42em width0em depth1.42em} \right) 
\ea
\lab{bfmA.3}
\ee 
\normalsize
where $\tilde{A}=-1.171953$. Note that in the main part of the text we used notation
$\mu^2\rightarrow 4\pi\mu^2$.
\bb{99}
\bibitem{hatfield} T. Barnes  and G. I.  Chandour , Phys. Rev. D {\bf 22} 924 (1980 )
\nwl
S. J. Chang, Phys. Rev. {\bf 12}, 1071, (1975);\\
B. Hatfield, "{\it Quantum Field Theory of Point Particles and Strings}",
(Addison - Wesley, Reading, 1992).
\bibitem{amelino} G. Amelino - Camelia and So - Young Pi, Phys. Rev. D
 {\bf 47}, 2356, (1993).
\bibitem{okop} A. Okopinska Phys. Rev. D {\bf 35}, 1835, (1987) ;\\
A. Okopinska, "{\it Optimized Expansion}". BK21 Tutorial lectures.
 Institute of Physycs \& Applied Physics,
Yonsei University, Seoul, 2001.
\bibitem{stev42} I. Stancu and P. M. Stevenson,  Phys. Rev. D {\bf 42}, 2710, (1990).
\bibitem{cea55} P. Cea and L. Tedesco  Phys. Rev. D {\bf 55}, 4967, (1997).
\bibitem{yee} G.H. Lee and J.H. Yee,  Phys. Rev. D {\bf 56}, 6573,  (1997);\\
G. H.  Lee, T.H.  Lee and  J. H. Yee,  
Phys.Rev. D {\bf 58}, 125001, (1998). 
\bibitem{abbot}B. S. DeWitt, Phys. Rev. 
{\bf 162}, 1195,  (1967);\\
 L. F. Abbott, "{\it Introduction to Back Ground Field Method}" , Lectures presented at the XXI
Cracow school of theoretical Physics, Ref. TH3113-CERN, 1981;\\
Chooky Lee, "{\it Selected Topics in Theoretical Physics}", Proc. of the Fifth Symposium on Theoretical Physics
Seoul, July, 1986, Edited by H. S. Song.
\bibitem{stev32}  P.M. Stevenson,  Phys. Rev. D {\bf  32}, 1389,  (1985).
\bibitem{ramosprd60} M. R. Pinto and R. Ramos, Phys. Rev. D60,105005, (1999)
\bibitem{chiku} S. Chiku and T. Hatsuda  Phys. Rev. D{\bf 58}, 076001, (1998); \\
S. Chiku Progr. Theor.
Phys. 104, 1129, (2000).
\bibitem{banerji} N. Banerjee and S. Malik Phys. Rev. D{\bf 43},3368, (1991).
\bibitem{itzykson} J. Iliopoulos, C. Itzykson and A. Martin, Rev. Mod. Phys. 47,
165, (1975).
%\bibitem{jones} H. F. Jones and P. Parkin, Nucl. Phys. {\bf B594}, (2001), 518
\bibitem{stev23}  P.M. Stevenson,  Phys. Rev. D {\bf  23}, 2916,  (1981).
\bibitem{ramos65} F. de Souza Cruz, M. B. Pinto, R. Ramos and P. Sena 
 Phys. Rev. A {\bf  65}, 053613,  (2002).
\bibitem{chung} J. M. Chung and B.K. Chung , Phys. Rev. D{\bf 56}, 6508, (1997);\\
A.I. Davydychev and J.B. Tausk  Phys. Rev. D {\bf  53}, 7381,  (1996).
\eb

 %$\fill{\diamond} $ $\suitdiamond$ $\fill{\suitdiamond} $
\newpage
\centerline {\bf FIGURE CAPTIONS}
\begin{description}
\item [Fig. 1(a).]
 BFM Feynman diagrams contributing to the
 effective potential in $\delta^2$  order  (see Eq. \re{bfm3.23}).
 Solid lines represent the propagator  $G(p)=1/( p^2 + \Omega^2 )$. 
The vertices      $v_2  =(m_{0}^{2} -\Omega_{0}^{2})/2+6\lambda_0\phi^2 $,  $   v_3  =4\lambda_0\phi$  
and $    v_4= \lambda_0$ are marked by diamonds, circles and squares, respectively.

\item [Fig.1(b).]
 Feynman diagrams contributing to the effective potential in $\delta^2$
 order in the method of refs. \ci{okop,stev42}  (see Eq. \re{bfm4.1}). 
 These are obtained by an attachment of the ring diagram, $I_0$ , 
 denoted here by the small circle, to the each  two body vertex $v_2$  
of  the FIG.1(a). Solid lines represent the propagator $G(p)=1/( p^2 + \Omega^2 )$.
The vertices      $u_2  =(m_{0}^{2} -\Omega_{0}^{2})/2+6\lambda_0\phi^2 $,
  $   v_3  =4\lambda_0\phi$,   
 $    v_4= \lambda_0$  and   $    u_4= 6\lambda_0$ are marked by diamonds, circles,  filled   and opened 
squares, respectively.
 \item [Fig.2(a).]
The same as in FIG.1(a). but in  $\delta^3$  order.                                                                                    
\item [Fig.2(b).]
 Feynman diagrams contributing to the effective potential in $\delta^3$  order of ref. \ci{okop}. 
 These are obtained by attachment of a ring diagram, $I_0$ ,
 denoted here by the small circle, to the each  two body vertex $v_2$  of  the Fig.2(a).
 The notations are the same as in FIG.1(b).                                                           
\item [Fig.3.]
The same as in FIG.1(a) but in $\delta^4$  order. Note that, only those, including quadratic
 vertex $v_2$ are presented.  Implementation of a ring diagram to each $ v_2$  
vertex with all possible ways will give 18 additional  diagrams. 
 \end{description}
%%%%%%%%%%%%%%%%%%%%%%%%%%%%%%%%%%%%%%%%%%%%
%\newpage
\begin{figure}
 \epsfclipon
 \epsfxsize=12.cm
\begin{center}
\epsffile{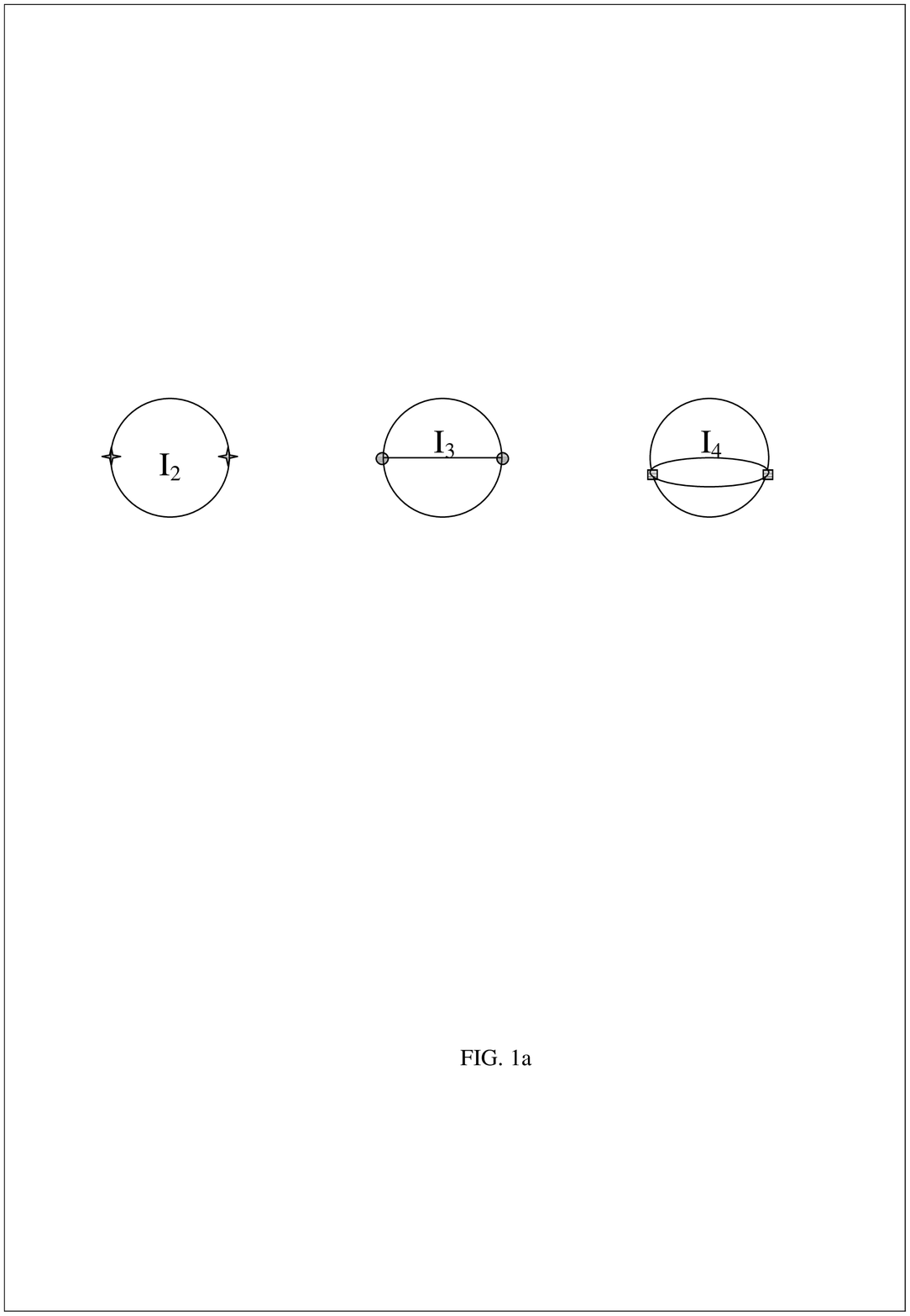}
 \end{center}
%\caption{\label{Fig.1a}BFM Feynman diagrams contributing to the
 %effective potential in $\delta^2$  order  (see Eq. \re{bfm3.23}).
 %Solid lines represent the propagator  $G(p)=1/( p^2 + \Omega^2 )$. 
%The vertices      $v_2  =(m_{0}^{2} -\Omega_{0}^{2})/2+6\lambda_0\phi^2 $,  
%$   v_3  =4\lambda_0\phi$  
%and $    v_4= \lambda_0$ are marked by diamonds, circles and squares, respectively.}
\end{figure}
%%%%%%%%%%%%%%%%%%%%%%%%%%%%%%%%%%%%%%%%%%%
\begin{figure}[ht]
 \epsfclipon
 \epsfxsize=12.cm
 \begin{center}
{\epsffile{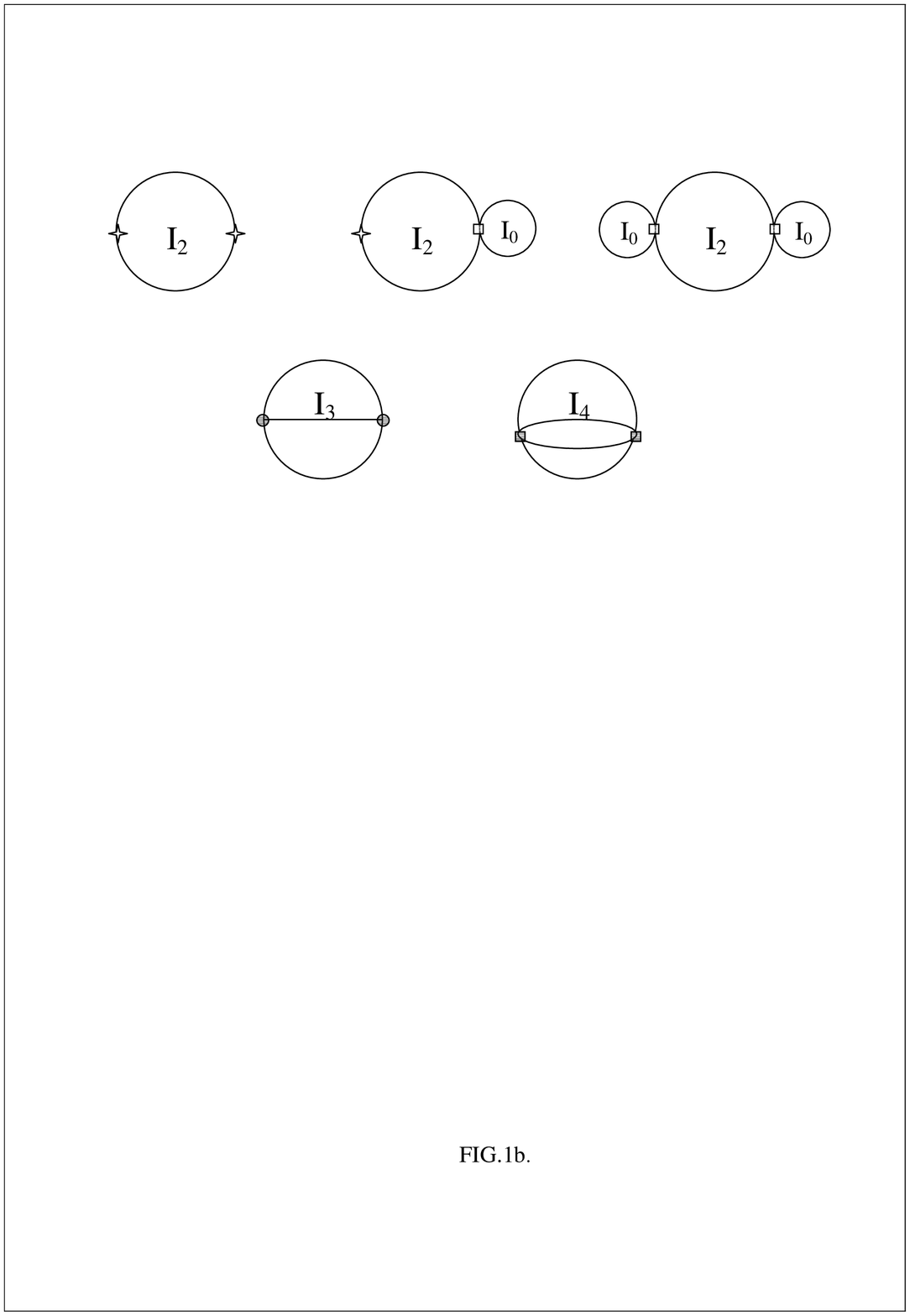}}
 \end{center}
%\caption{ Feynman diagrams contributing to the effective potential in $\delta^2$
%order in the method 
%of refs.  \cite{stev42}}
%  (see Eq. \re{bfm4.1}). }
 %These are obtained by an attachment of the ring diagram, $I_0$ , 
 %denoted here by the small circle, to the each  two body vertex $v_2$  
%of  the FIG.1(a). Solid lines represent the propagator $G(p)=1/( p^2 + \Omega^2 )$.
%The vertices      $u_2  =(m_{0}^{2} -\Omega_{0}^{2})/2+6\lambda_0\phi^2 $,
 % $   v_3  =4\lambda_0\phi$,   
 %$    v_4= \lambda_0$  and   $    u_4= 6\lambda_0$ are marked 
%by diamonds, circles,  filled   and opened} 
%squares, respectively.} 
\end{figure}
%%%%%%%%%%%%%%%%%%%%%%%%%%%%%%%%%%%%%%%%%%%
\begin{figure}[ht]
 \epsfclipon
 \epsfxsize=12.cm
 \centerline{\epsffile{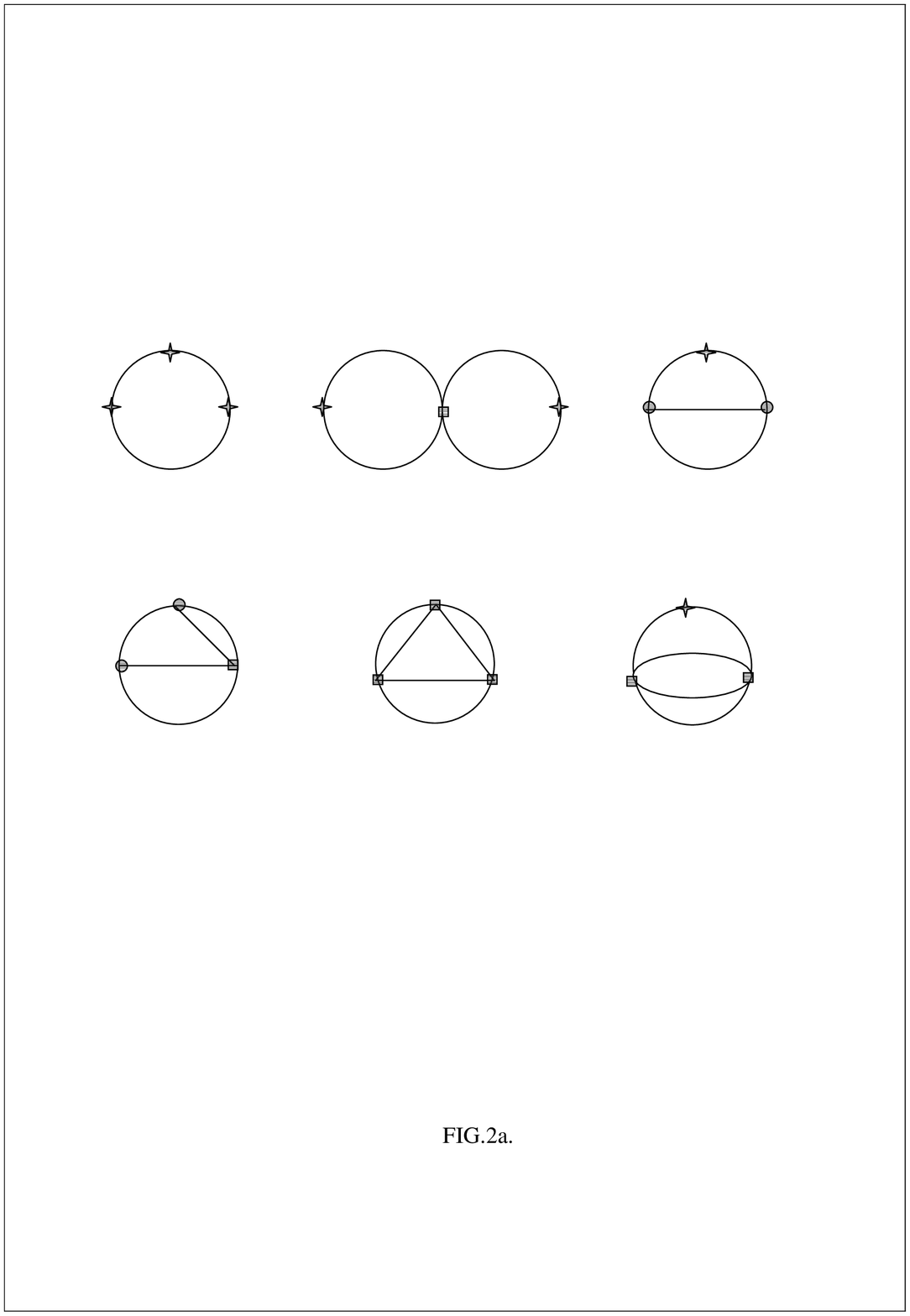}}
%\caption{\label{Fig.2a}The same as in FIG.1(a). but in  $\delta^3$  order.       }
 \end{figure}
%%%%%%%%%%%%%%%%%%%%%%%%%%%%%%%%%%%%%%%%%%%%%
\begin{figure}[ht]
 \epsfclipon
 \epsfxsize=12.cm
 \centerline{\epsffile{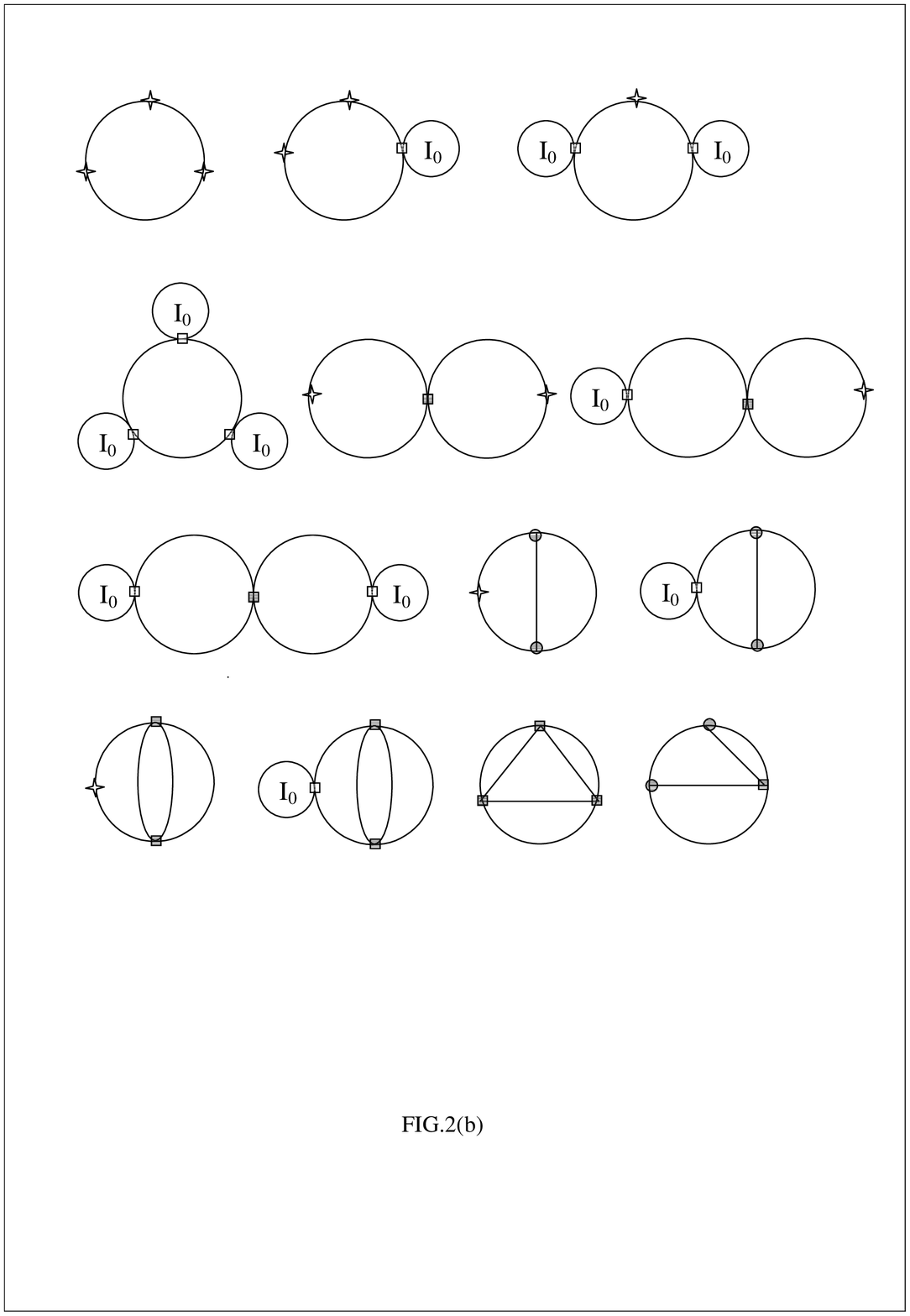}}
%\caption{\label{Fig.2b}   Feynman diagrams contributing to the effective potential in $\delta^3$  order of ref. \ci{okop}. 
 %These are obtained by attachment of a ring diagram, $I_0$ ,
 %denoted here by the small circle, to the each  two body vertex $v_2$  of  the Fig.2(a).
 %The notations are the same as in FIG.1(b).         }
 \end{figure}
%%%%%%%%%%%%%%%%%%%%%%%%%%%%%%%%%%%%%%%%%%
\begin{figure}[ht]
 \epsfclipon
 \epsfxsize=12.cm
 \centerline{\epsffile{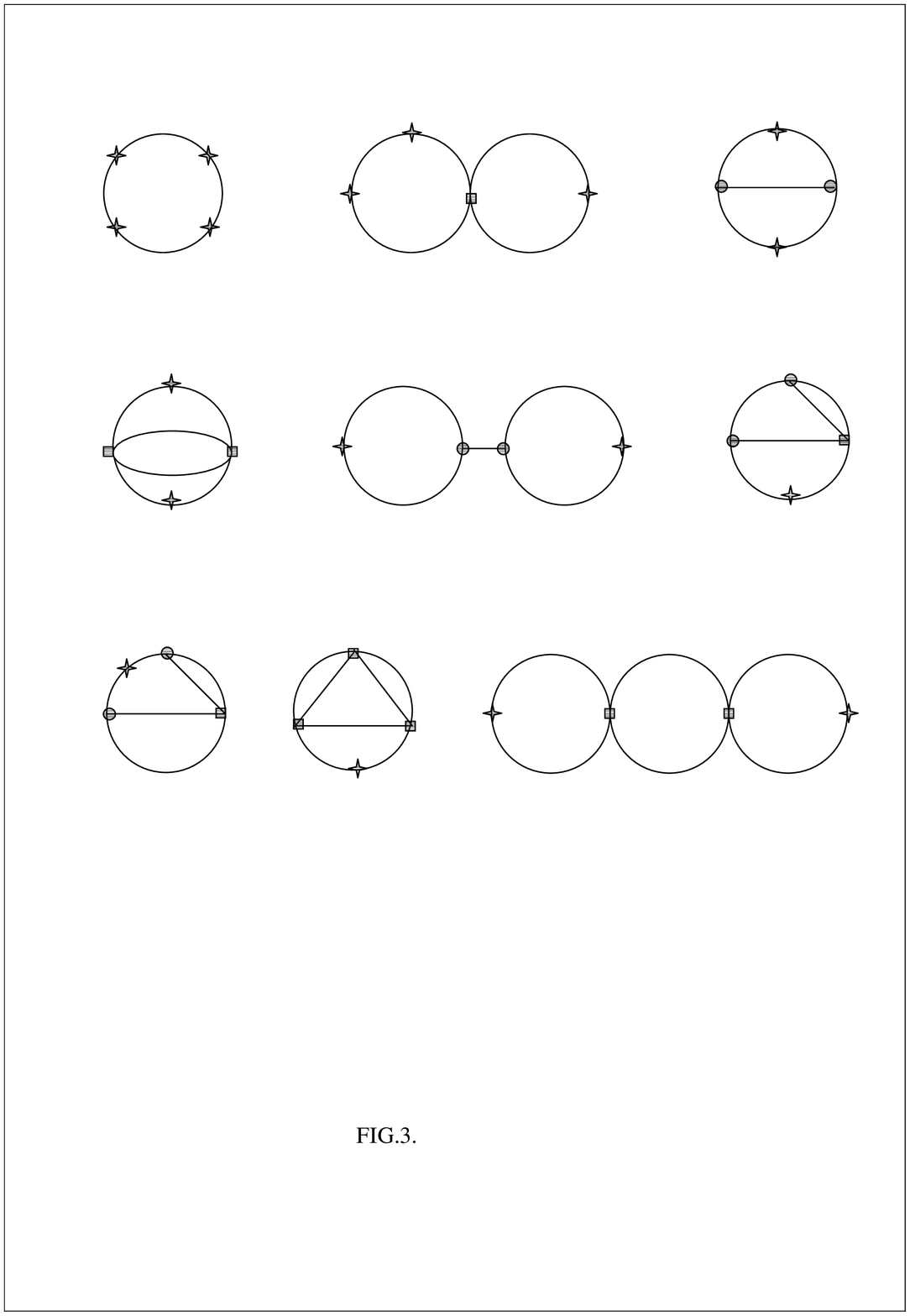}}
%\caption{\label{Fig.3}   The same as in FIG.1(a) but in $\delta^4$  order. Note that, only those, including quadratic
 %vertex $v_2$ are presented.  Implementation of a ring diagram to each $ v_2$  
%vertex with all possible ways will give 18 additional  diagrams.        }
 \end{figure}
%%%%%%%%%%%%%%%%%%%%%%%%%%%%%%%%%%%%%%%%%%
%%%%%%%%%%%%%%%%%%%%%%%%%%%%%%%%%%%%%%%%%%
\end{document}